\documentclass[12pt]{article}
\usepackage{ol2}

\usepackage{amsfonts,amssymb,amsmath,bm}

\begin{document}

%\twocolumn[

\title{Giant enhancement of electric field between two close
  metallic grains due to plasmonic resonance}

 \author{V.Lebedev$^{1,2}$, S.Vergeles$^{1,2}$ and P.Vorobev$^1$}
 \address{$^1$Landau Institute for Theoretical Physics RAS, Kosygina 2, 119334, Moscow, Russia}
 \address{$^2$Moscow Institute of Physics and Technology, Dolgoprudnyj, Institutskij
 per. 9, 141700, Moscow Region, Russia}

 \begin{abstract}
We theoretically examine plasmonic resonance between two close metallic grains separated
by a gap of width much less than the length of the incident electromagnetic wave.
Resonance conditions are established and the electric field enhancement is found. Our
general arguments are confirmed by analytic solution of the problem for simplest
geometries. We discuss an extension of our results to more complex cases.
 \end{abstract}


%]

Interaction of electromagnetic field with metallic dimers recently became a focus of
intensive research due to extraordinary potentials for various applications ranging from
nanophotonics to bio/chemical sensing. A key feature of such nanostructures utilized in
these applications is a significant field enhancement in the dimer's gap. This
enhancement is primarily studied experimentally or using computer modeling, which does
not allow us to uncover the relation between the characteristics of the near field
enhancement and the physical properties and geometry of the dimer and host material. In
this paper we present an analytical description of the characteristics of the field
enhancement between two metallic nanogranules embedded into the host dielectric.

Resent progress in nanofabrication have led to thriving activity in the actual design of
composite materials with plasmonic sub-wavelength dimensions for applications in
photonics and chemical sensing. They can be realized as surface grain structures on a
dielectric substrate, see, e.g., Refs. \cite{KreVol,Kneipp}. The giant electric field
enhancement is observed in narrow gaps between metallic grains. Strongly amplified
electromagnetic fields can be generated both in a broad spectral range for disordered
metal-dielectric composites and at selected frequency ranges for periodically ordered
nano-structures. The literature devoted to the problem is immense, for introduction to
the research region a reader can use the monograph \cite{07SS}, for brief examination of
the disordered composites see \cite{ZhangX06}.

The plasmonic resonances can be excited at propagation of an electromagnetic wave through
a composite material where metal grains are inserted into a dielectric matrix. It is
known, see Refs. \cite{77Boa,78Bho}, that the plasmonic resonance in a single metallic
grain is excited at a frequency near the plasma one lying in the ultraviolet spectral
region for ``good'' metals (Ag, Au, etc.). To reach the resonance in the optical or near
infrared diapason, one should use a special geometry where metallic grains are separated
by distances much smaller than their size. Then giant electric field enhancement occurs
in the gap between the grains at resonance conditions implying, particularly, large
negative permittivity of the grains. Here we theoretically investigate the phenomenon.
The resonance conditions are related mainly to the local geometrical characteristics of
the gap. The field enhancement, on the contrary, is determined by a variety of
geometrical factors controlling the energy flow to the gap. We present general arguments
that enable us to estimate both the resonance frequencies and the electric field
enhancement for grain dimers. The arguments are confirmed by analytic solution of the
problem for two identical metallic spheres. Then we extend our consideration to more
complicated cases. Applications of our results to periodic and disordered
metal-dielectric composites are also discussed.

We consider the case of a monochromatic electromagnetic wave of frequency $\omega$, then
the electric field strength is written as $\mathrm{Re}\, [\bm E \exp(-i\omega t)]$. The
permittivity of the matrix, $\varepsilon_\mathrm{d}(\omega)$, is assumed to be of order
unity, with small imaginary part. We accept a local relation between the electric field
strength and the electric displacement field $\bm D$, $\bm D=\varepsilon_\mathrm{m}
(\omega)\bm E$, in the metal grains. Here $\varepsilon_\mathrm{m}$ is the permittivity of
the metal. In optical and near infrared spectral regions the permittivity of a ``good''
metal can be described by Drude-Lorentz formula
 \begin{equation}
 \varepsilon_\mathrm{m}\sim
 -(\omega_\mathrm{p}/\omega)^2
 \left[1-i/(\omega\tau)\right],
 \label{Drude}
 \end{equation}
where $\omega_\mathrm{p}$ is the plasma frequency and $\tau$ is the electron relaxation
time. Therefore in the frequency interval $\omega_\mathrm{p}\gg\omega\gg \tau^{-1}$ the
permittivity $\varepsilon_\mathrm{m}$ has negative real part, large by its absolute
value, and relatively small imaginary part. The same is true for the dielectric contrast
$\varepsilon=\varepsilon_\mathrm{m}/\varepsilon_\mathrm{d}$. Thus, we arrive at the
condition necessary for the giant electric field enhancement.

We examine mainly the case where two close metallic grains are surrounded by an unbounded
dielectric medium. The grains are assumed to be small that is their sizes are much less
than the wavelength $\lambda$ of the electromagnetic wave (in the dielectric medium). We
are looking for the electromagnetic field profile near the grains, especially in the gap
between the grains, where one expects an essential enhancement of the field. The problem
belongs to the so-called near-field optics. The electric field near small metallic grains
can be examined in quasi-electrostatic approximation (see, e.g., Ref. \cite{07SS}), in
terms of its potential $\Phi$. We consider the case where the dielectric medium and the
metal have isotropic dielectric properties. Then the electric potential $\Phi$ satisfies
Laplace equation $\nabla^2 \Phi =0$ both in the dielectric medium and in the metallic
grains. The equation has to be supplemented by the boundary conditions at the
metal-dielectric interface, which are continuities of the electric field tangent
component and of the electric displacement field component normal to the interface.

We are interested in the electric field strength enhancement, characterized by the ratio
$E_c/E_0$, where $E_c$ is the electric field strength in the central segment of the gap
between the grains, and $E_0$ is the electric field strength in the incident
electromagnetic wave. An essential enhancement has to be observed near resonance
frequencies, then there are two principal contributions to the ratio,
 \begin{equation}
 {E_c}/{E_0}=
 {G}/(\varepsilon-\varepsilon_\mathrm{res}) + G_\mathrm{bg},
 \label{ratio}
 \end{equation}
that can be called resonance and background terms. The quantity
$\varepsilon_\mathrm{res}$, as well as the factors $G$ and $G_\mathrm{bg}$, are
determined by geometry of the grains and by the gap thickness. A maximum value of the
enhancement is observed at the resonance condition,
$\varepsilon'=\varepsilon_\mathrm{res}$, then $|E_c/E_0|\approx G/\varepsilon''$, where
$\varepsilon'$ and $\varepsilon''$ are real and imaginary parts of the dielectric
contrast $\varepsilon$. Thus, at the condition, the electric field enhancement is
restricted by energy losses. The resonance frequency can be evaluated now from the
expression (\ref{Drude}), $\omega_\mathrm{res}\sim \omega_\mathrm{p}/
\sqrt{|\varepsilon_\mathrm{res}|}$.

% \begin{figure}[tb]
% \includegraphics[scale=0.45]{./PlasmonShort_OL.eps} \
% \caption{a) Narrow gap between two metallic grains. b) Two parallel elongated grains.
% c) Two coaxial elongated grains.}
% \label{fig:gap}
% \end{figure}

\begin{figure}[h]
 \includegraphics[scale=0.45]{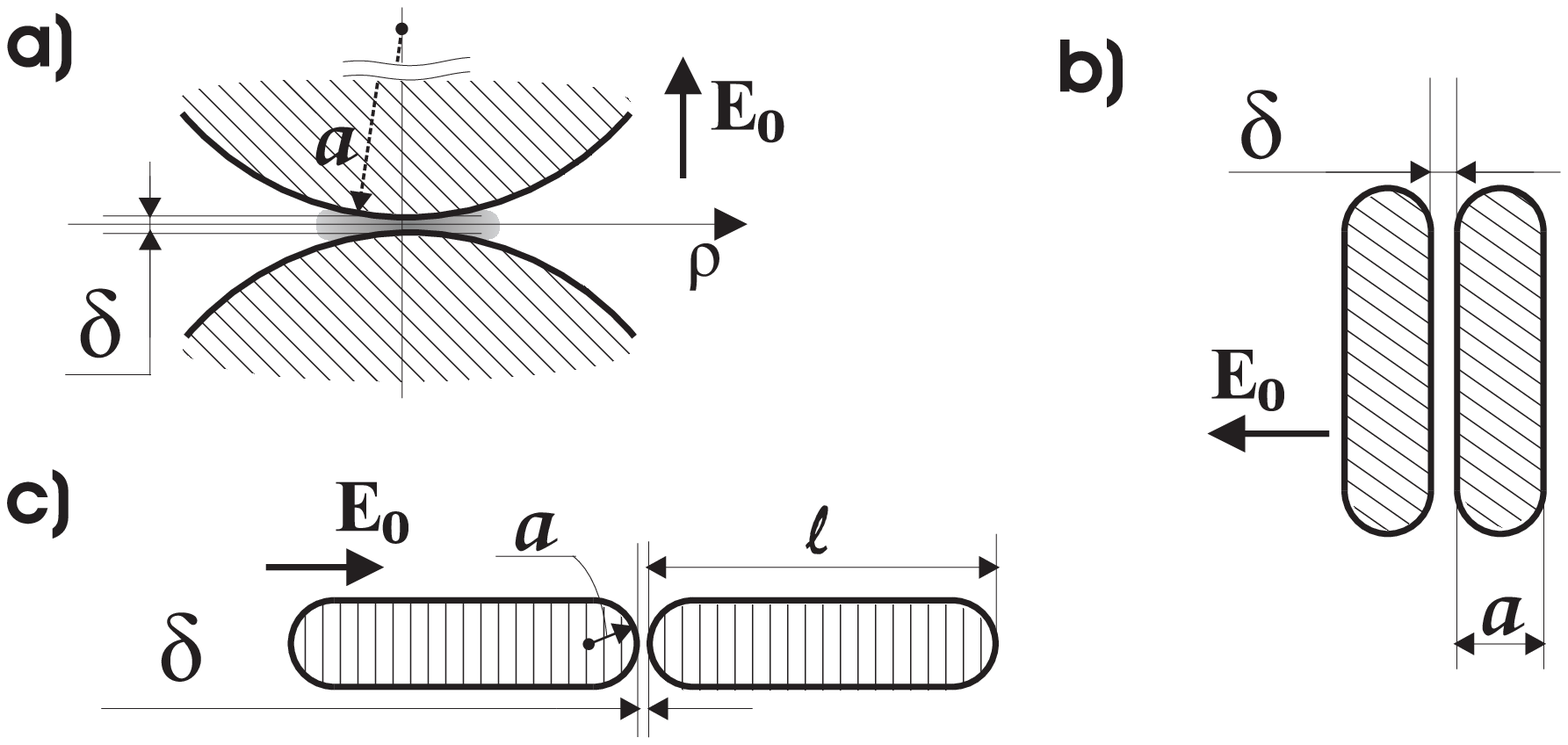} \
 \caption{a) Narrow gap between two metallic grains. b) Two parallel elongated grains.
 c) Two coaxial elongated grains.}
 \label{fig:gap}
 \end{figure}

In the case of close grains the electric field of the resonance mode is localized in the
central segment of the gap between the grains where it can be regarded as flat (grey
region in Fig. \ref{fig:gap}a). An electromagnetic wave can propagate along a narrow flat
gap between two metallic bodies separated by a dielectric medium provided $\varepsilon=
-\coth (\beta\delta/2)$ where $\beta$ is the propagation constant and $\delta$ is the gap
width \cite{74KM}. We are interested in the case $|\varepsilon|\gg1$, then the above
relation can be rewritten as $\beta=-2/(\varepsilon \delta)$. Standing waves are
determined by the conditions $\beta L \approx \pi n$ where $n$ is an integer number and
$L$ is the longitudinal size of the gap. For two close smooth grains the size of the flat
region between the grains can be estimated as $\sqrt{a\delta}$, where $a$ is
characteristic grain curvature radius, see Fig. \ref{fig:gap}a. Thus we obtain the
following estimation for the resonance values of the permittivity
 \begin{equation}
 \varepsilon_\mathrm{res}
 \sim -\sqrt{a/\delta}\big/(n+\delta n),
 \label{resonance}
 \end{equation}
where $n$ is an integer number and $\delta n\sim 1$. The estimation is valid at $n\ll
\sqrt{a/\delta}$, that ensures the condition $\beta\delta\ll1$ (it is equivalent to the
inequality $|\varepsilon|\gg1$).

The electric field inside the gap is approximately homogeneous at distances
$\rho\lesssim\sqrt{a\delta}$ from the gap center where the gap can be regarded as flat.
At distances $a\gg\rho\gg\sqrt{a\delta}$ from the gap center the gap thickness is
estimated as $\rho^2/a$ and is much larger than in the center. The potential difference
between the grain boundaries is $\rho$-independent in the main approximation, since
$\varepsilon$ is effectively infinite at the scales. The condition is equivalent to the
law $E\propto \rho^{-2}$. Therefore we arrive at the estimation $E\sim E_c
a\delta/\rho^2$. The field strength determines the charge density at the grain boundaries
that scales $\propto \rho^{-2}$ as well. Therefore the dipole moment $d$ of the grains is
determined by the distances $\sim a$ and can be estimated as $d\sim E_c \delta a^2$.

The geometrical factor $G$ in Eq. (\ref{ratio}) can be evaluated from balancing between
Ohmic dissipation rate $Q$ and the energy pumping. At the resonance frequency, the
dissipation rate is estimated as
 \begin{equation}
 Q\sim\omega \varepsilon''(\varepsilon')^{-2}
 E_c^2 (a\delta)^{3/2},
 \label{losses}
 \end{equation}
where the last factor represents the metal volume where the dissipation occurs. The
energy pumping can be estimated as $\omega d E_0\sim \omega E_c a^2\delta E_0$. Comparing
the expressions one finds $G\sim(a/\delta)^{3/2}$, that is $G\gg 1$ at our conditions.
One can also evaluate the dipole radiation intensity
 \begin{equation}
 I\sim \omega E_c^2 \delta^2 a^4/\lambda^3.
 \label{radiation}
 \end{equation}
The radiation leads to additional energy losses, our scheme is correct provided $I\ll Q$.

The presented qualitative picture is confirmed by rigorous analytical calculations for
the case where the metallic grains are close identical spheres of radii $a$ separated by
a distance $\delta\ll a$. Our problem can be solved analytically by passing to the
so-called bispherical reference system \cite{64Dav,76GN}. Here, we do not discuss details
of the calculations (they will be published elsewhere) and present final results only.
With logarithmic accuracy, one finds for polarization of the incident wave parallel to
the symmetry axis
 \begin{eqnarray}
 \varepsilon_\mathrm{res}=-\frac{\sqrt{a/\delta}}{n+1/2}, \quad
 G= \frac{8\pi^2}{3\ln(a/\delta)},
 \nonumber \\
 G_\mathrm{bg} = -2\sqrt{a/\delta}, \quad
 d=\frac{a^2\delta E_c}{3\ln(a/\delta)},
 \label{final}
 \end{eqnarray}
where $n=0,1,\dots$. The expressions (\ref{final}) are in accordance with our above
reasoning.

One can test more sophisticated grain geometries that are characterized by a number of
scales. We examine strongly prolate grains of length $l$, their mutual arrangement can be
either parallel or coaxial (see. Fig. \ref{fig:gap}b,c) with a narrow gap $\delta$
between them. The resonance condition is determined by the smallest longitudinal gap
dimension. Hence, one can use the same relation (\ref{resonance}), where $a$ is the
smallest curvature radius of the grains characterizing the gap. It is the cylinder radius
in the parallel case and curvature radius of the grain tips in the coaxial case. The
field enhancement $G$ in Eq. (\ref{ratio}) should be found, as above, from the energy
balance: in the case of parallel cylinders $G\sim a/\delta$, and in the case of elongated
grains $G\sim l a^{1/2} \delta^{-3/2}$.

In our analysis, we used the quasi-static approximation that implies that all the
characteristic sizes of the grains are much less than the electromagnetic wavelength
$\lambda$ in the homogeneous dielectric medium. Besides, we established that the
resonance mode is localized between the grains, at distances $\rho\lesssim
\sqrt{a\delta}$ from the gap center. And just the vicinity determines the resonance
condition (\ref{resonance}). Therefore the condition survives provided the quasistatic
approximation does work at the scale, that is if the scale $\sqrt{a\delta}$ is less than
$\lambda/\sqrt{|\varepsilon|}$ where $\lambda$ is the wave length in dielectric. That
leads to the condition $\lambda^2\gg a^{3/2}\delta^{1/2}$, justifying Eq.
(\ref{resonance}) even for large grains. However, the amplification factor in this case
should be determined using complete geometry of the system and Maxwell equations.

Since the resonance mode is localized between the grains, a principal role in giant
electric field enhancement for a random distribution of the grains is played by grain
dimers with suitable separations, satisfying the resonance condition. Thus a number of
sharp peaks in the space distribution of the electric field has to be observed in the
disordered metal—dielectric composite in the external electromagnetic wave. Experimental
data \cite{ZhangX03,Jain} qualitatively proves the conclusion, see, e.g., Refs.
\cite{Shalaev,Markel,02Shal}. To determine the number of peaks at a given frequency one
has to know a probability distribution of small separations $\delta$ (in comparison with
the grain sizes) in grain dimers.

Recently considerable efforts are applied in designing periodic metal-dielectric
composites, see, e.g., Ref. \cite{Drachev3}. One could imagine a periodic structure of
metallic grains (say, of metallic spheres) separated by small distances. In this case
resonance modes can be excited where the electric field has sharp maxima in central
segments of the gaps between the grains. However, due to overlapping of the modes
localized near the gaps, the resonance has to be transformed into a band of delocalized
modes, like it occurs for electrons in a periodic potential (crystalline lattice). As our
analysis shows, the structure of the electric field in the gap between the grains at
distances smaller than the grain size, is fixed by boundary conditions at the
metal-dielectric interface. Therefore the resonance frequency has to be determined by
matching conditions in the regions between the grains, instead of the condition at
infinity for two grains immersed in an unrestricted dielectric medium. Thus we expect
that the band width is of the order of the separation between the resonance frequencies.
Thus, relying on a power-like dependence like in Eq. (\ref{Drude}), we conclude that for
grains characterized by a single size $a$ the band width is of the order of the resonance
frequency itself.

One of our assumptions was smoothness of the grain boundaries. If the boundaries are
rough then a problem appears concerning an enhancement of energy losses observed in Ref.
\cite{08Drachev}. The giant electrical field enhancement lead to increasing non-linear
effects. The problems need a special investigation and are out of the scope of this work.

We thank I. Gabitov and A. Sarychev for numerous valuable discussions. The work is partly
supported by Russian Federal Education Agency, contract P587.

\end{document}